\documentclass[12pt,usenames,dvipsnames,a4paper]{article}

\usepackage{soul}

\usepackage[utf8]{inputenc}
\usepackage[T1]{fontenc}

\usepackage{changepage}
\usepackage{amsmath,amsfonts,amssymb}
\usepackage{epsf,amsmath,bbold,amsfonts,stmaryrd}
\usepackage{mathrsfs}
\usepackage{appendix}
\usepackage{caption}
\usepackage{color}
\usepackage{datetime}
\usepackage{float}
\usepackage{graphicx}
\usepackage[colorlinks,hyperindex,breaklinks]{hyperref}
\hypersetup{pageanchor=false,citecolor=red,urlcolor=red}
\usepackage{indentfirst}
\usepackage[numbers,square,comma,sort&compress,merge]{natbib} 
\usepackage{subcaption}
\usepackage{mathtools}
\usepackage{ytableau}
\usepackage{tabu}
\usepackage{esvect}

\usepackage{calc}
\usepackage{bm}

\allowdisplaybreaks

\def\c{\chi}

\def\d{\delta}

\def\eps{\varepsilon}
\def\f{\frac}

\def\G{\Gamma}

\def\l{\left}

\def\mc{\mathcal}

\def\m{\mu}
\def\n{\nu}

\def\p{\partial}

\def\r{\right}
\def\s{\sigma}
\def\t{\tau}

\def\be{\begin{equation}}
\def\ee{\end{equation}}

\def\bea{\begin{eqnarray}}
\def\eea{\end{eqnarray}}

\def\ba{\begin{array}}
\def\ea{\end{array}}

\def\bc{\begin{center}}
\def\ec{\end{center}}

\def\bl{\begin{flushleft}}
\def\el{\end{flushleft}}

\def\br{\begin{flushright}}
\def\er{\end{flushright}}

\def\bi{\begin{itemize}}
\def\ei{\end{itemize}}

\def\bt{\begin{tabular}}
\def\et{\end{tabular}}

\makeatletter

\newsavebox\myboxA
\newsavebox\myboxB
\newlength\mylenA

\newcommand*\xoverline[2][0.75]{%
    \sbox{\myboxA}{$\m@th#2$}%
    \setbox\myboxB\null
    \ht\myboxB=\ht\myboxA%
    \dp\myboxB=\dp\myboxA%
    \wd\myboxB=#1\wd\myboxA
    \sbox\myboxB{$\m@th\overline{\copy\myboxB}$}
    \setlength\mylenA{\the\wd\myboxA}
    \addtolength\mylenA{-\the\wd\myboxB}%
    \ifdim\wd\myboxB<\wd\myboxA%
       \rlap{\hskip 0.5\mylenA\usebox\myboxB}{\usebox\myboxA}%
    \else
        \hskip -0.5\mylenA\rlap{\usebox\myboxA}
         {\hskip 0.5\mylenA\usebox\myboxB}%
    \fi}
\makeatother

\def\be{\begin{equation}}
\def\ee{\end{equation}}

\def\bea{\begin{eqnarray}}
\def\eea{\end{eqnarray}}

\def\f{\frac}

\def\p{\partial}

\usepackage{xspace}

\usepackage{xifthen}
\newcommand*\diff{\mathrm{d}} 
\newcommand*\ldiff[2][]{ \ifthenelse{\isempty{#1}}{ \diff
#2}{\diff^#1#2} \,} 
\let\limitint\int 
\renewcommand{\int}{\limitint \!} 

\begin{document}

\begin{titlepage}
\vspace{5cm}

\vspace{2cm}

\begin{center}
     \Large\textbf{Geometrical origin of inflation\\ 
     in Weyl-invariant Einstein-Cartan gravity}
\end{center}

\vspace{1cm}

\begin{center}
{\textsc {Georgios K. Karananas}}
\end{center}

\begin{center}
{\it Arnold Sommerfeld Center\\
Ludwig-Maximilians-Universit\"at M\"unchen\\
Theresienstra{\ss}e 37, 80333 M\"unchen, Germany}\\

\end{center}

\begin{center}
\texttt{\small georgios.karananas@physik.uni-muenchen.de} \\
\end{center}

\vspace{2cm}

\begin{abstract}

It is shown that the scalar degree of freedom built-in in the quadratic
Weyl-invariant Einstein-Cartan gravity can drive inflation and with
predictions in excellent agreement with observations.

\end{abstract}

\end{titlepage}

Ref.~\cite{Karananas:2024xja} constructed the unique, ghost-free,
Weyl-invariant quadratic gravity in the Einstein-Cartan-Sciama-Kibble
(EC) formulation of General Relativity\,\footnote{We work from the onset in
the ``affine picture'' with variables the metric $g_{\m\n}$ and affine
connection $\G^\m_{~\n\rho}$. These are related to the gauge fields of
translations~(tetrad $e^A_\m$) and Lorentz transformations~(spin connection
$\omega^{AB}_\m$) as
\begin{equation*}
g_{\m\n} = e^A_\m \eta_{AB} e^B_\n \ ,~~~\G^\m_{~\n\rho} = e^\m_A \l(\p_\n e^A_\rho+\omega^A_{\n B}e^B_\rho\r) \ ,
\end{equation*} 
where $\eta_{AB}$ is the Minkowski metric and capital Latin letters stand for
Lorentz indexes. }
\be
\label{eq:infl_action_full}
S = \int \diff^4 x \sqrt{g} \Bigg[ \f{1}{f^2}R^2 +\f{1}{\tilde f^2}\tilde R^2 +\f{1}{\tilde g^2}R\tilde R \Bigg] \ .
\ee 
Here $f,\tilde f$ and $\tilde g$ are gauge couplings of the Lorentz group,
$g=-{\rm det}(g_{\m\n})$, and
\be
\label{eq:scalar_curvatures}
R = g^{\sigma\n} \d^\m_\rho R^\rho_{~\sigma\m\n} \ ,~~~\tilde R = E^{\rho\s\m\n}R_{\rho\s\m\n} \ ,
\ee
with $E^{\m\n\rho\s}=\f{\eps^{\m\n\rho\s}}{\sqrt{g}}$, are the scalar and
pseudoscalar (Holst) curvatures built out of the affine curvature tensor
\be
\label{eq:curvature_tensor}
R^\rho_{~\sigma\m\n} = \p_\m  \G^\rho_{~\n\s} - \p_\n \G^\rho_{~\m\sigma} + \G^\rho_{~\m\lambda} \G^\lambda_{~\n\sigma} -\G^\rho_{~\n\lambda} \G^\lambda_{~\m\sigma} \ ,
\ee 
where $\G^\m_{~\n\rho}$ is the torsionful, metric-compatible, affine
connection. Note that one can even drop the requirement of metricity and work
in the context of the full-blown metric-affine gravity(MAG)---our conclusions
are the same.

As far as the dynamics is concerned, the action~(\ref{eq:infl_action_full}) is
classically equivalent to General Relativity (with non-zero cosmological
constant related to $f$), supplemented by a massive spin-0 field minimally
coupled to gravity and with a non-trivial potential, owing to the presence of
the $R\tilde R$ term. As it will become clear, it is exactly because of this
that the scalar can play the role of the inflaton.

We now obtain the equivalent theory in the purely metrical description by
following the procedure of Ref.~\cite{Karananas:2024xja}: 

\emph{i)}~We start by bringing the action~(\ref{eq:infl_action_full}) into its
 ``first-order form'' by introducing two auxiliary fields, the
 spurion/dilaton $\c$ and scalar $\phi$\,\footnote{There is no unique way to
 express an action in terms of auxiliary fields; for instance, instead of
     (\ref{eq:first_order_action}), we could have equally well rewritten~(\ref
 {eq:infl_action_full}) as
\be
\label{eq:equivalent_first_order_action}
S = \int\diff^4 x\sqrt{g}\Bigg[\c^2 R + M_P^2\phi\tilde R -\f{\tilde f^2 \l(2q\c^2-M_P^2\phi\r)^2}{4\l(1-4q\tilde q\r)} -\f{f^2\c^4}{4}\Bigg] \ ,
\ee
or even as~\cite{Gialamas:2024iyu}
\be
\label{eq:another_equivalent_first_order_action}
S=\int\diff^4 x\sqrt{g}\Bigg[\c^2 R + M_P^2\phi\tilde R - \f{\tilde g^2}{{1-4q\tilde q}}\l(q\c^4-4q\tilde q M_P^2\c^2\phi +\tilde q M_P^4\phi^2\r)\Bigg] \ .
\ee 
Notice that~(\ref{eq:another_equivalent_first_order_action}) boils down to
(\ref{eq:equivalent_first_order_action}) by ``completing the square'' via the
addition and subtraction of $f^2\c^4/4$. In turn, shifting $\phi$ to
$\phi+2q\c^2/M_P^2$ in the action~(\ref
{eq:equivalent_first_order_action}) gives~(\ref
{eq:first_order_action}), which is our starting point. }
\be
\label{eq:first_order_action}
S = \int\diff^4 x\sqrt{g}\Bigg[\c^2 R + \l(2q\c^2+M_P^2\phi\r)\tilde R -\f{\tilde f^2 M_P^4\phi^2}{4\l(1-4q\tilde q\r)} -\f{f^2\c^4}{4}\Bigg] \ ,
\ee 
where as obvious from the above, we took $\c$ with mass-dimension one and
$\phi$ with mass-dimension zero\,\footnote{The assignment of dimensions is
completely arbitrary, so we chose the most convenient one.}---we also
introduced
\be
q = \f{f^2}{4\tilde g^2} \ ,~~~\tilde q = \f{\tilde f^2}{4\tilde g^2} \ .
\ee
It can be easily checked that on the equations of motion for $\c$ and $\phi$,
the above coincides with~(\ref{eq:infl_action_full}).  

\emph{ii)}~The Weyl invariance of the theory allows for the convenient gauge
 choice
\be
\label{eq:Weyl_gauge}
\c = \f{M_P}{\sqrt 2} \ ,
\ee
and the first-order action~(\ref{eq:first_order_action}) becomes 
\be
\label{eq:action_weyl_gauge}
S = M_P^2\int\diff^4 x\sqrt{g}\Bigg[\f R 2 + \l(q+\phi\r)\tilde R -\f{\tilde f^2 M_P^2 \phi^2}{4\l(1-4q\tilde q\r)}-\f{M_P^2f^2}{16}\Bigg] \ .
\ee

\emph{iii)}~Next, we split the connection into the Levi-Civita part
 $\mathring \G^{\m}_{~\n\rho}$ plus torsional contributions (see e.g.~\cite
 {Diakonov:2011fs,Rasanen:2018ihz,Karananas:2021zkl,Rigouzzo:2022yan})
\be
\label{eq:connection_decomp}
\G^{\m}_{~\n\rho} = \mathring \G^{\m}_{~\n\rho}  +\f 1 3\l(g_{\n\rho}v_\m - \d^\m_\n v_\rho\r) +\f{1}{12} E^\m_{~\n\rho\s}a^\s - \tau_{\n\rho}^{~~\m} \ ,
\ee
where $v_\m,a_\m,\tau_{\m\n\rho}$ are the usual irreducible pieces of the
torsion tensor $T^\m_{~\n\rho}\equiv \G^\m_{~\n\rho}-\G^\m_
{~\rho\n}$, defined as
\be
v^\m = g_{\n\rho}T^{\n\m\rho} \ ,~~~a^\m = E^{\m\n\rho\s}T_{\n\rho\s} \ ,~~~\tau_{\m\n\rho} = T_{\m\n\rho}  +\f 1 3\l(g_{\m\n}v_\rho-g_{\m\rho}v_\n \r) - \f 1 6 E_{\m\n\rho\s}a^\s \ ,
\ee
with $\tau_{\m\n\rho}=-\tau_{\m\rho\n},~g^{\m\rho}\tau_{\m\n\rho}=E^
{\m\n\rho\s}\tau_{\n\rho\s} = 0$.  Using~(\ref{eq:connection_decomp}) and the
expressions~(\ref{eq:curvature_tensor},\ref{eq:scalar_curvatures}), we find
that the scalar and Holst curvatures are decomposed as 
\begin{align}
\label{eq:R_resol}
&R = \mathring R + 2\mathring \nabla_\m v^\m -\f 2 3 v_\m v^\m + \f{1}{24} a_\m a^\m +\f 1 2 \t_{\m\n\rho} \t^{\m\n\rho} \ ,\\
\label{eq:tildeR_resol}
&\tilde R = -\mathring \nabla_\m a^\m +\f 2 3 a_\m v^\m +\f 1 2 E^{\m\n\rho\s} \t_{\lambda\m\n}\t^\lambda_{\ \rho\s} \ ,
\end{align} 
with $\mathring R$ the (Riemannian) Ricci scalar and $\mathring\nabla_\m$ the
torsion-free covariant derivative. 

\emph{iv)}~We then plug~(\ref{eq:R_resol},\ref{eq:tildeR_resol}) into~(\ref
 {eq:action_weyl_gauge}) and after dropping full divergences, we end up with
\begin{align}
\label{eq:action_resolved_curvatures}
S = M_P^2&\int\diff^4 x\sqrt{g}\Bigg[\f{\mathring R}{2} +\f 1 4 \t_{\m\n\rho} \t^{\m\n\rho} +\f{q+\phi}{2}E^{\m\n\rho\s} \t_{\lambda\m\n}\t^\lambda_{\ \rho\s}\nonumber\\
& -\f 1 3 v_\m v^\m + \f{2\l(q+\phi\r)}{3}a_\m v^\m +\f{1}{48}a_\m a^\m -\phi\mathring\nabla_\m a^\m  -\f{\tilde f^2 M_P^2 \phi^2}{4\l(1-4q\tilde q\r)}-\f{M_P^2f^2}{16}\Bigg] \ .
\end{align}

\emph{v)}~We now vary the above wrt $v_\m, a_\m$ and $\tau_
 {\m\n\rho}$,  resulting into the following algebraic equations for torsion
\be
\label{eq:torsion_EOM}
v_\m -(q+\phi)a_\m = 0 \ ,~~~(q+\phi)v_\m +\f{1}{16}a_\m +\f 3 2 \p_\m\phi = 0 \ ,~~~\tau_{\m\n\rho}+2(q+\phi)E_{\kappa\lambda\n\rho}\tau_{\m}^{~\kappa\lambda} = 0 \ ,
\ee
from which we find
\be
\label{eq:torsion_shell}
v_\m = (q+\phi)a_\m \ ,~~~a_\m = -\f{24\p_\m\phi}{1+16(q+\phi)^2} \ ,~~~\tau_{\m\n\rho} = 0 \ .
\ee

\emph{vi)}~The penultimate step consists in using~(\ref{eq:torsion_shell}) to
 take the action~(\ref{eq:action_resolved_curvatures}) on-shell; this yields
\be
\label{eq:action_on-shell}
S = M_P^2\int\diff^4 x\sqrt{g}\Bigg[\f{\mathring R}{2} -\f{12}{1+16\l(q+\phi\r)^2}(\p_\m\phi)^2 -\f{\tilde f^2 M_P^2 \phi^2}{4\l(1-4q\tilde q\r)}-\f{M_P^2f^2}{16}\Bigg] \ .
\ee

\emph{vii)}~Finally, we make the kinetic term of $\phi$ canonical by
 introducing 
\be
\sqrt{\f 2 3} \f{\Phi}{M_P} =  {\rm arcsinh}\l[4(q+\phi)\r] \ ,
\ee
and~(\ref{eq:action_on-shell}) becomes
\be
\label{eq:final_action}
S = \int\diff^4 x\sqrt{g}\Bigg[\f{M_P^2}{2}\mathring R -\f 1 2(\p_\m\Phi)^2 - V(\Phi) -\f{M_P^4f^2}{16}\Bigg] \ ,
\ee
with 
\be
\label{eq:pseudo_potential}
V(\Phi) = V_0 \l(4q - \sinh\l[{\rm arcsinh}(4q)-\sqrt{\f 2 3 }\f{\Phi}{M_P}\r]\r)^2 \ ,~~~V_0 = \f{\tilde f^2 M_P^4}{64(1-4q\tilde q)} \ ,
\ee 
and we trivially shifted the field such that the minimum of its potential is
located at $\Phi=0$.

\begin{figure}[hbt!]
\centering
\begin{subfigure}[b]{.49\textwidth}
\centering
\includegraphics[width=\textwidth]{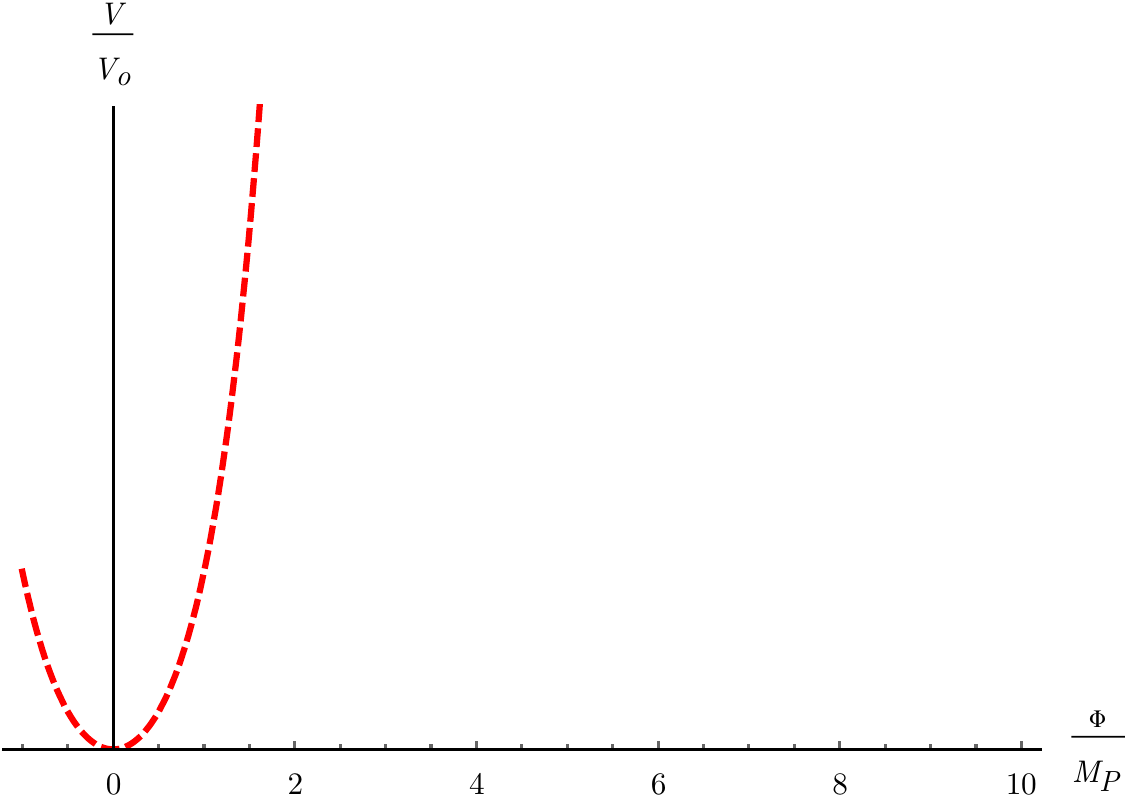}
\caption{$q=0$}
\end{subfigure}
\begin{subfigure}[b]{.49\textwidth}
\includegraphics[width=\textwidth]{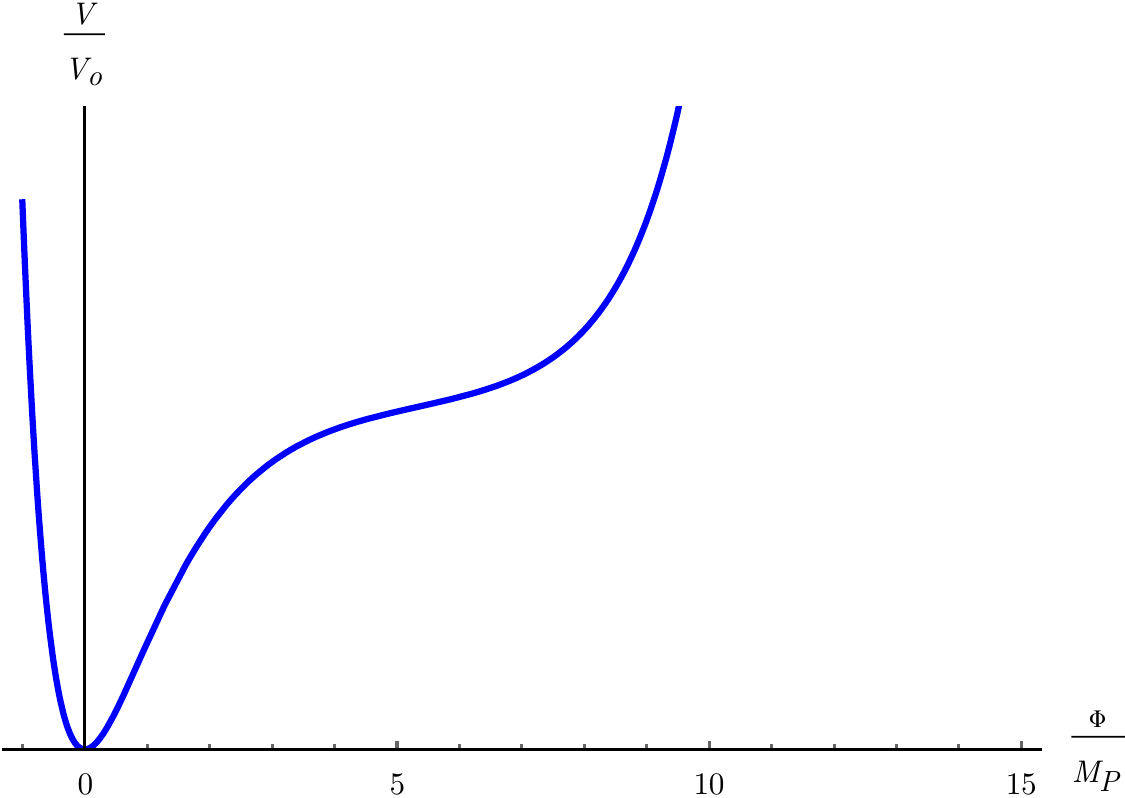}
\caption{$q=10$}
\end{subfigure}
\vskip\baselineskip
\begin{subfigure}[b]{.49\textwidth}
\centering
\includegraphics[width=\textwidth]{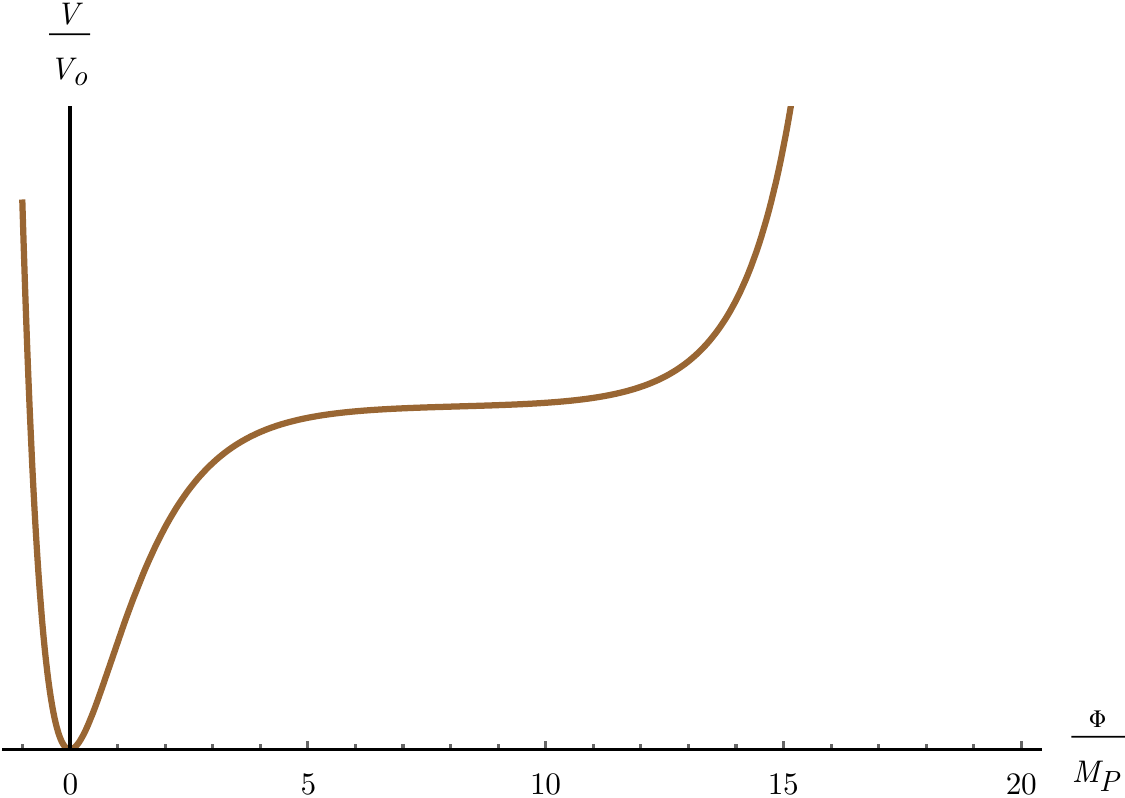}
\caption{$q=10^2$}
\end{subfigure}
\hfill
\begin{subfigure}[b]{.49\textwidth}
\includegraphics[width=\textwidth]{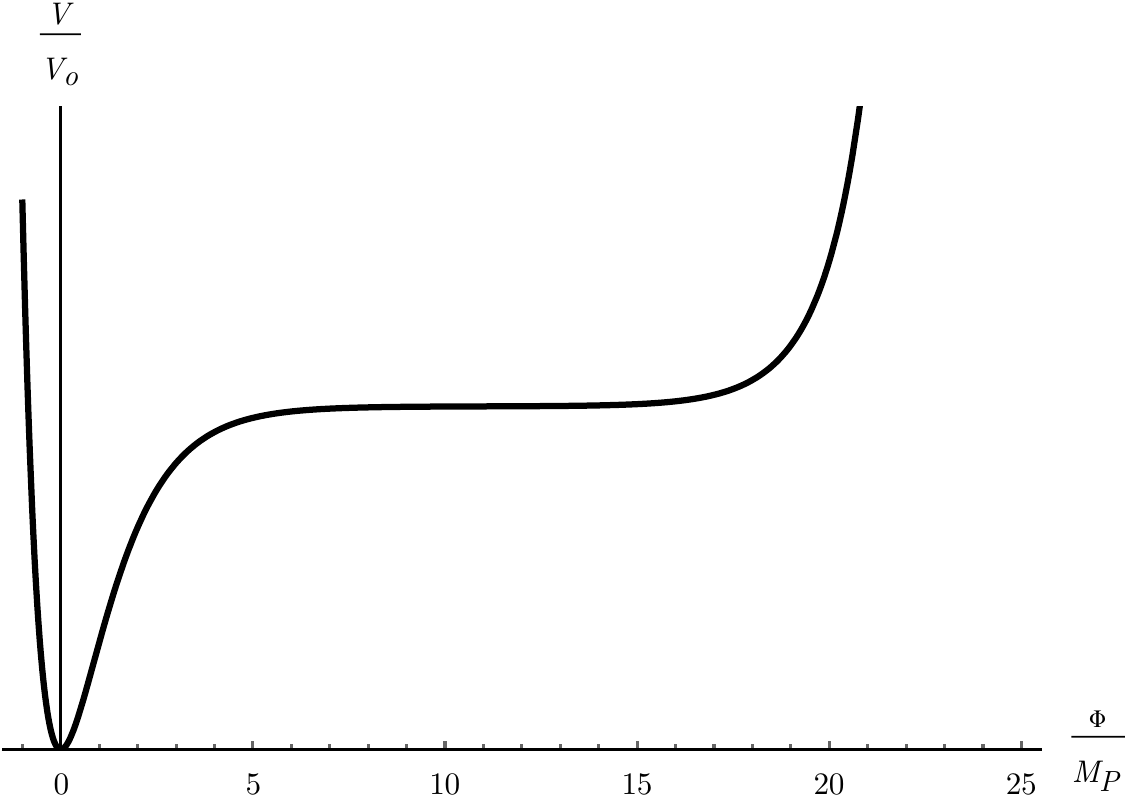}
\caption{$q=10^3$}
\end{subfigure}
\hfill
\caption{Plot of the potential~(\ref{eq:pseudo_potential}) for
 various values of $q$.}
\label{fig:ALP_plateau}
\end{figure}

Remarkably, the scalar field dynamics in the equivalent metric picture~(\ref
{eq:final_action},\ref{eq:pseudo_potential}) is identical to the one of
``pseudoscalaron inflation''~\cite
{Salvio:2022suk,Salvio:2023dba,DiMarco:2023ncs}; which has also been
obtained~(in MAG) from a different viewpoint, by requiring invariance under
``extended projective symmetry''~\cite{Barker:2024dhb}. 

This may be somewhat surprising at first sight, given that a term linear in
the Holst curvature, such that the potential for $q\gg 1$ have a
plateau~\cite
{Salvio:2022suk,Gialamas:2022xtt,Salvio:2023dba,DiMarco:2023ncs}, cannot
appear in the gravitational action~(\ref{eq:infl_action_full}) due to the
Weyl symmetry. Nevertheless, in the current setup and for all practical
purposes, $R\tilde R$ is the Holst term in disguise, aftermath of the fact
that the scalar curvature in~(scalar-curvature)$^2$ gravities is always
nonvanishing~\cite{Karananas:2024hoh,*Karananas:2024qrz}. 

The importance of the term that mixes $R$ and $\tilde R$ can also be
understood visually by inspecting Fig.~\ref{fig:ALP_plateau}, where $V
(\Phi)$ is plotted for various values of $q$. In its absence, i.e. for
$\tilde g\mapsto\infty$ or equivalently $q\mapsto 0$, the potential is too
steep to yield viable inflationary dynamics. As $q$ increases, the potential
exhibits a plateau and for (practically all) $q\gg 1$ the theory is capable
of accommodating ``good'' exponential expansion at early times. 

In passing, we note that our proposal (like pseudoscalaron inflation) bears a
conceptual resemblance to the Starobinsky $\mathring R + \mathring R^2$
model~\cite{Starobinsky:1980te}, in that the inflaton descends directly from
geometry. However, it should be stressed that if one wishes to stick to
metric gravity, the term linear in the Ricci scalar is absolutely essential.
Indeed, in the pure $\mathring R^2$ model---which is the metrical counterpart
of~(\ref{eq:infl_action_full})---the scalar spectrum comprises a massless
field (dilaton), see e.g.~\cite
{Alvarez-Gaume:2015rwa,Shtanov:2023lci,Karananas:2024hoh}, so it does not
allow for a graceful exit. Therefore,~\emph{inflation of purely geometrical
origin in~(scalar curvature)$^{\it 2}$~gravity necessitates its
Einstein-Cartan~(or metric-affine)~formulation.}

The analysis of inflation with the action~(\ref{eq:final_action}) is fairly
standard and has been worked out in details in~\cite
{Salvio:2022suk,Salvio:2023dba,DiMarco:2023ncs}, but for the sake of
completeness we also perform it now. The slow-roll parameters are given by 
\be
\eps = \f{M_P^2}{2}\l(\f{dV/d\Phi}{V}\r)^2 \ ,~~~\eta = M_P^2 \f{d^2V/d\Phi^2}{V} \ ,
\ee
which for~(\ref{eq:pseudo_potential}) become
\begin{align}
\label{eq:slow_roll_parameters}
&\eps = \f 4 3 \l(\f{\cosh\l[{\rm arcsinh}(4q)-\sqrt{\f 2 3 }\f{\Phi}{M_P}\r]}{4q -\sinh\l[{\rm arcsinh}(4q)-\sqrt{\f 2 3 }\f{\Phi}{M_P}\r]}\r)^2 \ ,\\
&\eta = \f 4 3 \f{\cosh\l[2{\rm arcsinh}(4q)-\sqrt{\f 8 3 }\f{\Phi}{M_P}\r]-4q \sinh\l[{\rm arcsinh}(4q)-\sqrt{\f 2 3 }\f{\Phi}{M_P}\r]}{\l(4q -\sinh\l[{\rm arcsinh}(4q)-\sqrt{\f 2 3 }\f{\Phi}{M_P}\r]\r)^2} \ .
\end{align}

Sufficient amount of exponential expansion requires that $\eps,|\eta|\ll 1$,
and the slow-roll approximation breaks down for $\eps\simeq 1$ or
$|\eta|\simeq 1$. For the case at hand, it is the $\eps$ parameter that
dictates when inflation ends, corresponding to\,\footnote{There exists yet
another real solution to $\eps\simeq 1$, which however lies outside the
inflationary domain.}
\be
\label{eq:sr_breakdown}
\sqrt{\f 2 3}\f{\Phi_{\rm end}}{M_P} ={\rm arcsinh}(4q)+\log\l[(2+\sqrt{3})\l(4\sqrt 3 q -\sqrt{\l(4\sqrt 3 q\r)^2-1}\r)\r] \ .
\ee

The number of inflationary efoldings between horizon exit $\Phi_\ast$ and the
end of inflation $\Phi_{\rm end}$ is 
\begin{align}
N &= \f{1}{M_P}\limitint^{\Phi_\ast}_{\Phi_{\rm e}}  \f{\diff \Phi}{\sqrt{2\eps}} \nonumber\\
&=\f 3 4 \log\l[\f{\cosh\l[{\rm arcsinh}(4q)-\sqrt{\f 2 3 }\f{\Phi_\ast}{M_P}\r]}{\sqrt 3 \l(8q-\sqrt{\l(4 \sqrt 3 q\r)^2-1}\r)}\r] \nonumber\\
& -3q\arctan\l[\f{\sinh\l[{\rm arcsinh}(4q)-\sqrt{\f 2 3 }\f{\Phi_\ast}{M_P}\r]+2\l(6q-\sqrt{\l(4\sqrt 3 q\r)^2-1}\r)}{1-2\l(6q-\sqrt{\l(4\sqrt 3 q\r)^2-1}\r)\sinh\l[{\rm arcsinh}(4q)-\sqrt{\f 2 3 }\f{\Phi_\ast}{M_P}\r]}\r] \ ,
\end{align}
where we used~(\ref{eq:sr_breakdown}).

To continue analytically and get a (rough) qualitative picture, for what
follows we take $q\gg 1$. We can then neglect the logarithm---it can be
checked that this is a good approximation~\cite{DiMarco:2023ncs} already for
$q> \mc O(10)$ and field values relevant for inflation---so that the above
can be inverted to give 
\be
\label{eq:field_horizon_exit}
\sqrt{\f 2 3}\f{\Phi_\ast}{M_P} \simeq {\rm arcsinh}(4q)+{\rm arcsinh}\Bigg[\f{2\sqrt 3-3 \cos\l(\f{2N}{3q}\r)-6q\sin\l(\f{2N}{3q}\r)}{6q\l(1-\cos\l(\f{2N}{3q}\r)\r)+3 \sin\l(\f{2N}{3q}\r)}\Bigg] \ .
\ee

\begin{figure}[t!]
\centering
\begin{subfigure}[b]{0.49\textwidth}
\centering
\includegraphics[width=\textwidth]{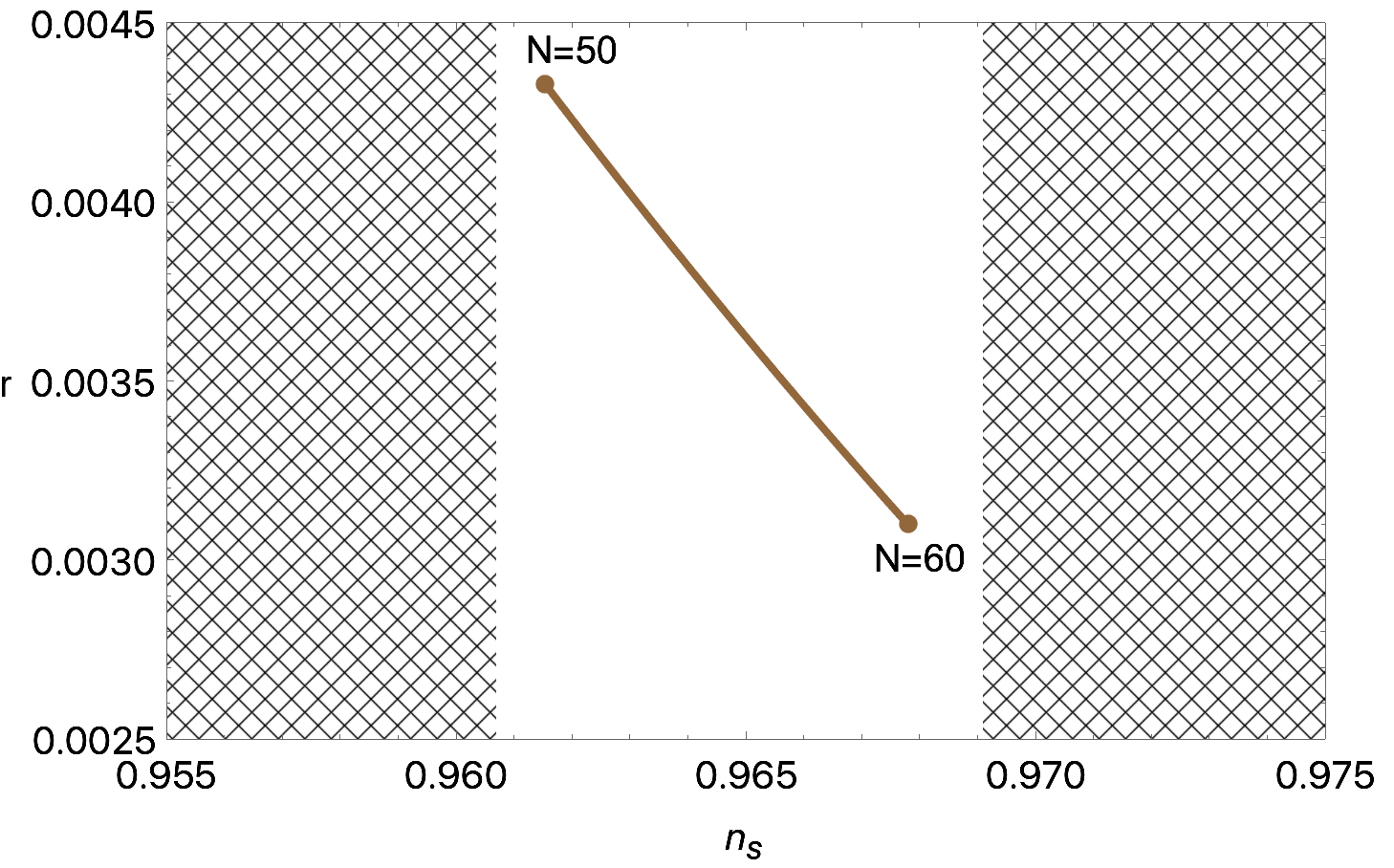}
\caption{$q=10^2$}
\end{subfigure}
\begin{subfigure}[b]{0.49\textwidth}
\centering
\includegraphics[width=\textwidth]{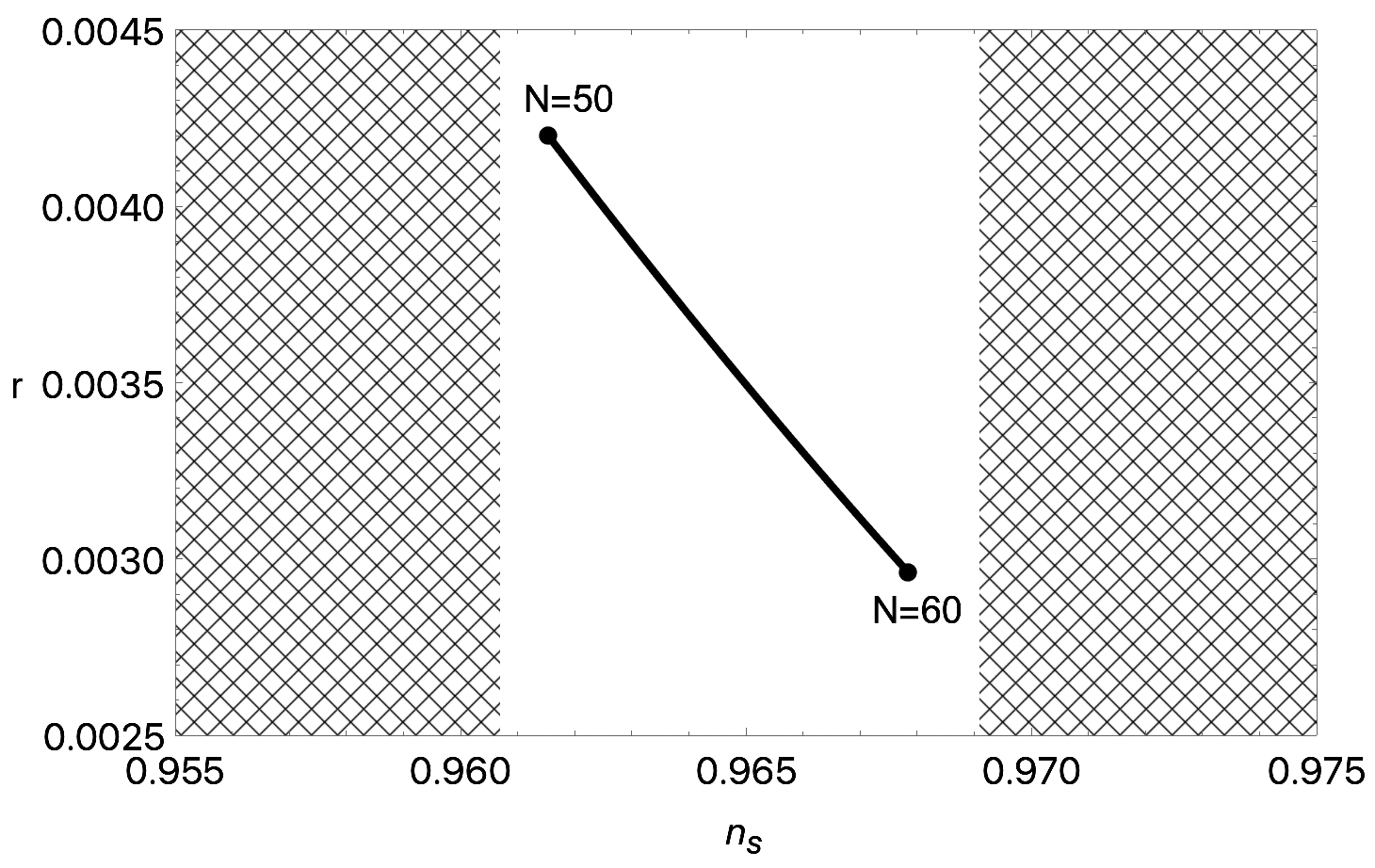}
\caption{$q=10^3$}
\end{subfigure}
\hfill
\caption{The predictions of the model in the usual $n_s-r$ plane for $q=10^2$
 (left) and $q=10^3$ (right). The observationally excluded regions at $1\s$
 for the spectral index~\cite{Planck:2018jri}, $n_s^{\rm obs}<0.9607$ and
 $n_s^{\rm obs}> 0.9691$, have been crossed out. The tensor-to-scalar ratio,
 being $r\sim\mc O(10^{-3})$, is comfortably below the upper bound $r^
 {\rm obs}<0.036~{\rm at}~2\s$~\cite{BICEP:2021xfz} (not shown in these
 plots). Since the observables are controlled by $q$ (for fixed $N$), the CMB
 normalization can be matched with appropriate $\tilde f$ given in~(\ref
 {eq:ftilde_CMB}), see Fig.~\ref{fig:ns_r_vs_q}.}
\label{fig:ns_vs_r}
\end{figure}

The Cosmic Microwave Background~\cite{Planck:2018jri} normalization dictates
that for $\Phi=\Phi_\ast$ 
\be 
\f{V}{\eps}  = 5\times 10^{-7} M_P^4 \ .
\ee 
Assuming for instance that $\tilde q\ll \f{1}{4q}$, so that the strength of
the potential $V_0$ (see Eq.~(\ref{eq:pseudo_potential})) is proportional to
$\tilde f^2 M_P^4$, we find
\be
\label{eq:ftilde_CMB}
\tilde f \simeq 7\times 10^{-3} \f{\sqrt{1+\Bigg(\f{2\sqrt 3-3 \cos\l(\f{2N}{3q}\r)-6q\sin\l(\f{2N}{3q}\r)}{6q\l(1-\cos\l(\f{2N}{3q}\r)\r)+3 \sin\l(\f{2N}{3q}\r)}\Bigg)^2}}{\l(4q+\f{2\sqrt 3-3 \cos\l(\f{2N}{3q}\r)-6q\sin\l(\f{2N}{3q}\r)}{6q\l(1-\cos\l(\f{2N}{3q}\r)\r)+3 \sin\l(\f{2N}{3q}\r)}\r)^2} \ .
\ee

Simply to get an estimate for the observables, let us fix e.g. $q\sim \mc O\l
(10^{3}\r)$ and $N\sim \mc O(60)$. Then 
\be
\tilde f \sim \mc O (10^{-8}) \ ,
\ee
from which the inflaton mass
\be
m_\Phi \simeq \f{\tilde f q M_P}{\sqrt 3} \ ,
\ee
is found to be in the ballpark of the scalaron mass in Starobinsky's
model~\cite{Starobinsky:1980te}, i.e.
\be
m_\Phi \sim \mc O\l(10^{-6}\r) M_P \ .
\ee

\begin{figure}[!t]
\centering
\begin{subfigure}[b]{\textwidth}
\includegraphics[width=.49\textwidth]{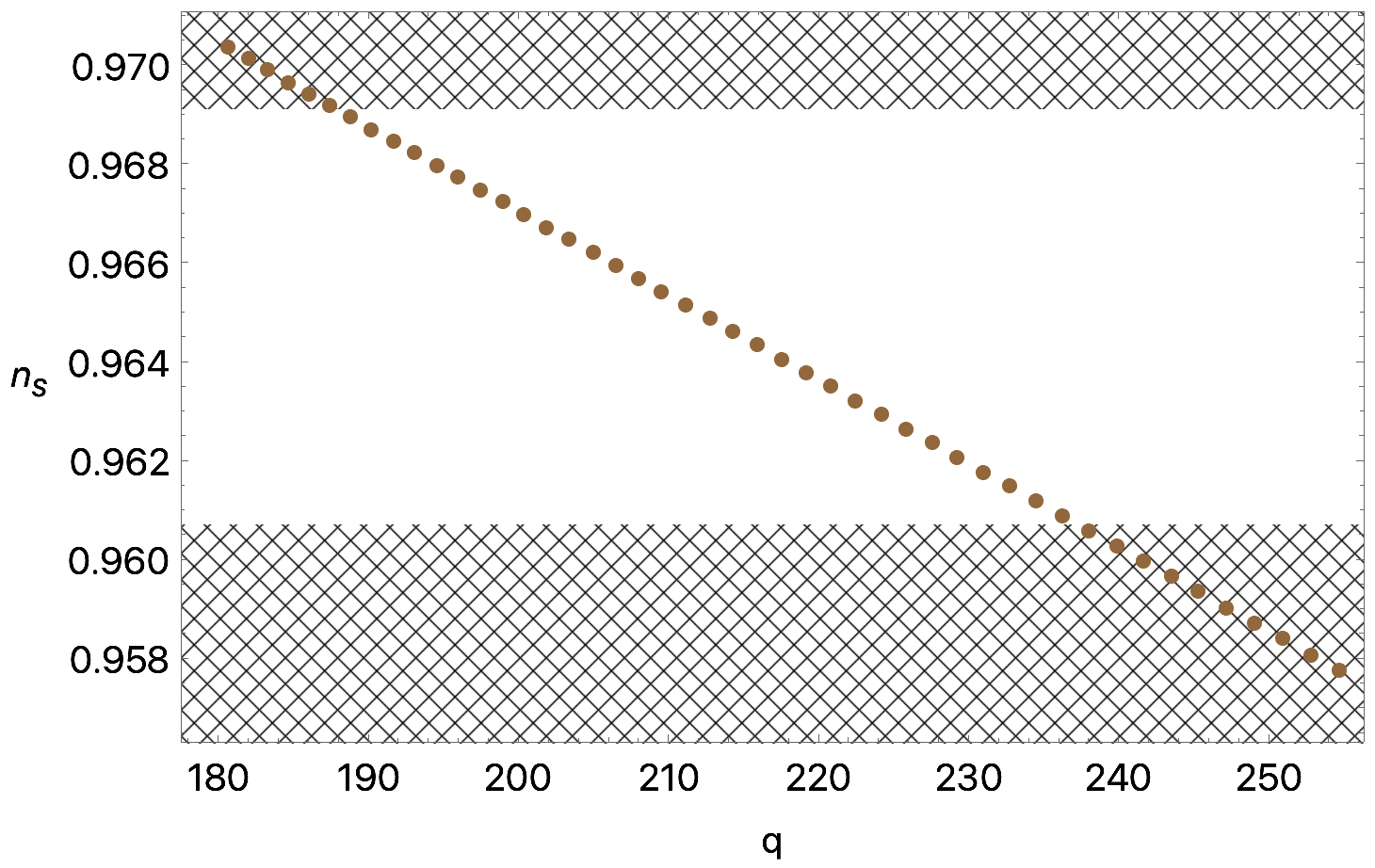}
\hfill
\includegraphics[width=.49\textwidth]{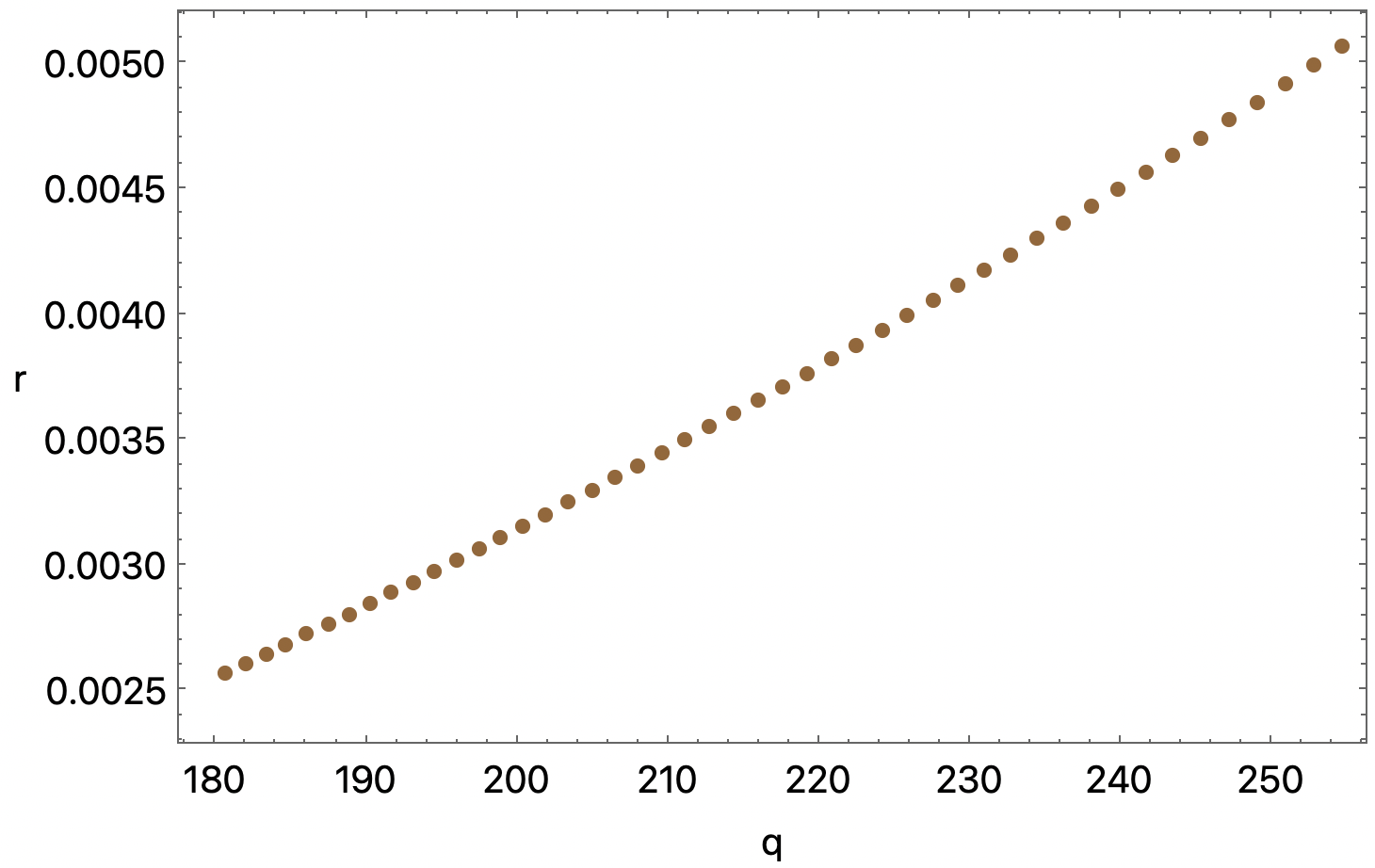}
\caption{$\tilde f=10^{-7}$}
\end{subfigure}
\vskip\baselineskip
\begin{subfigure}[b]{\textwidth}
\includegraphics[width=.49\textwidth]{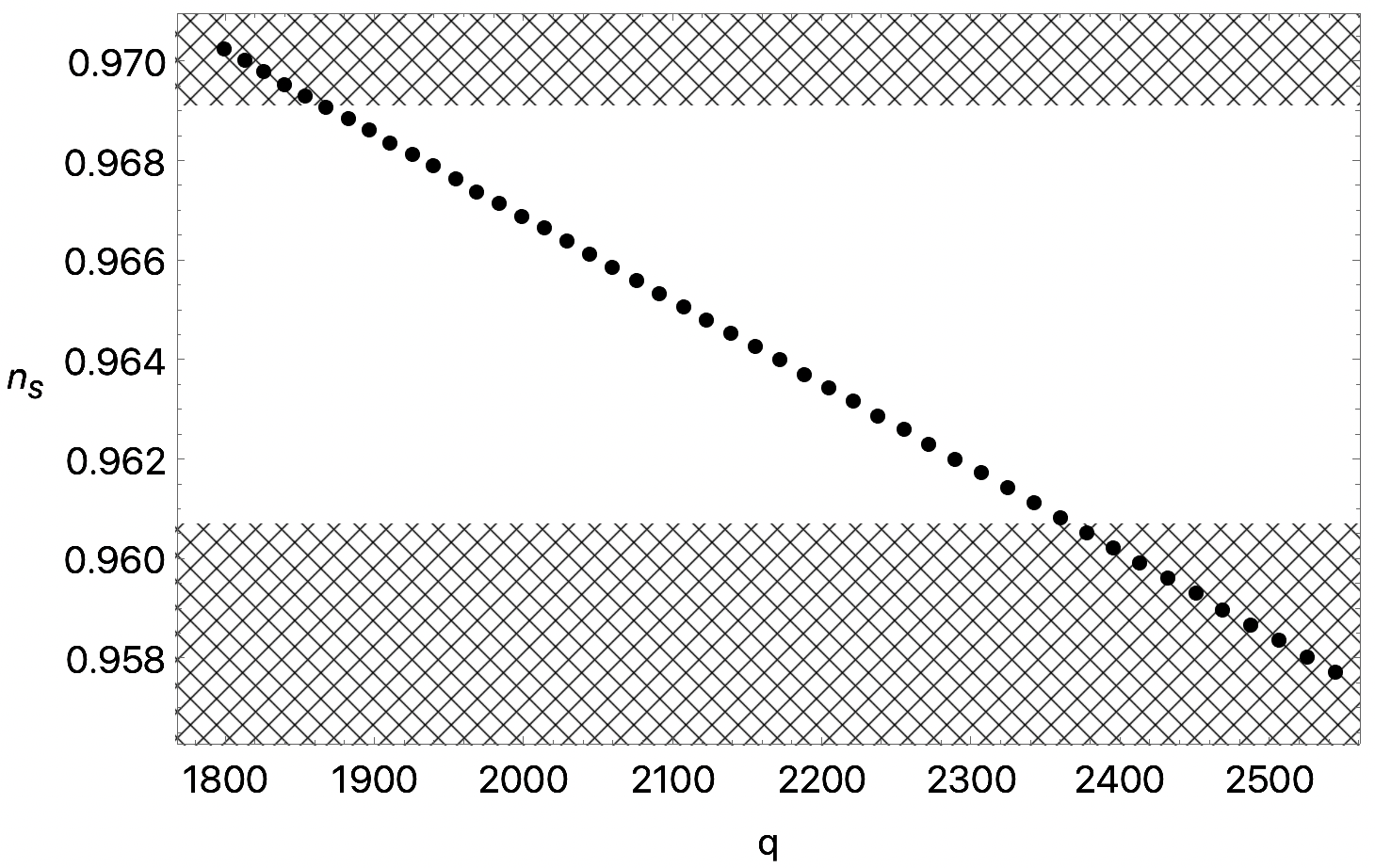}
\hfill
\includegraphics[width=.49\textwidth]{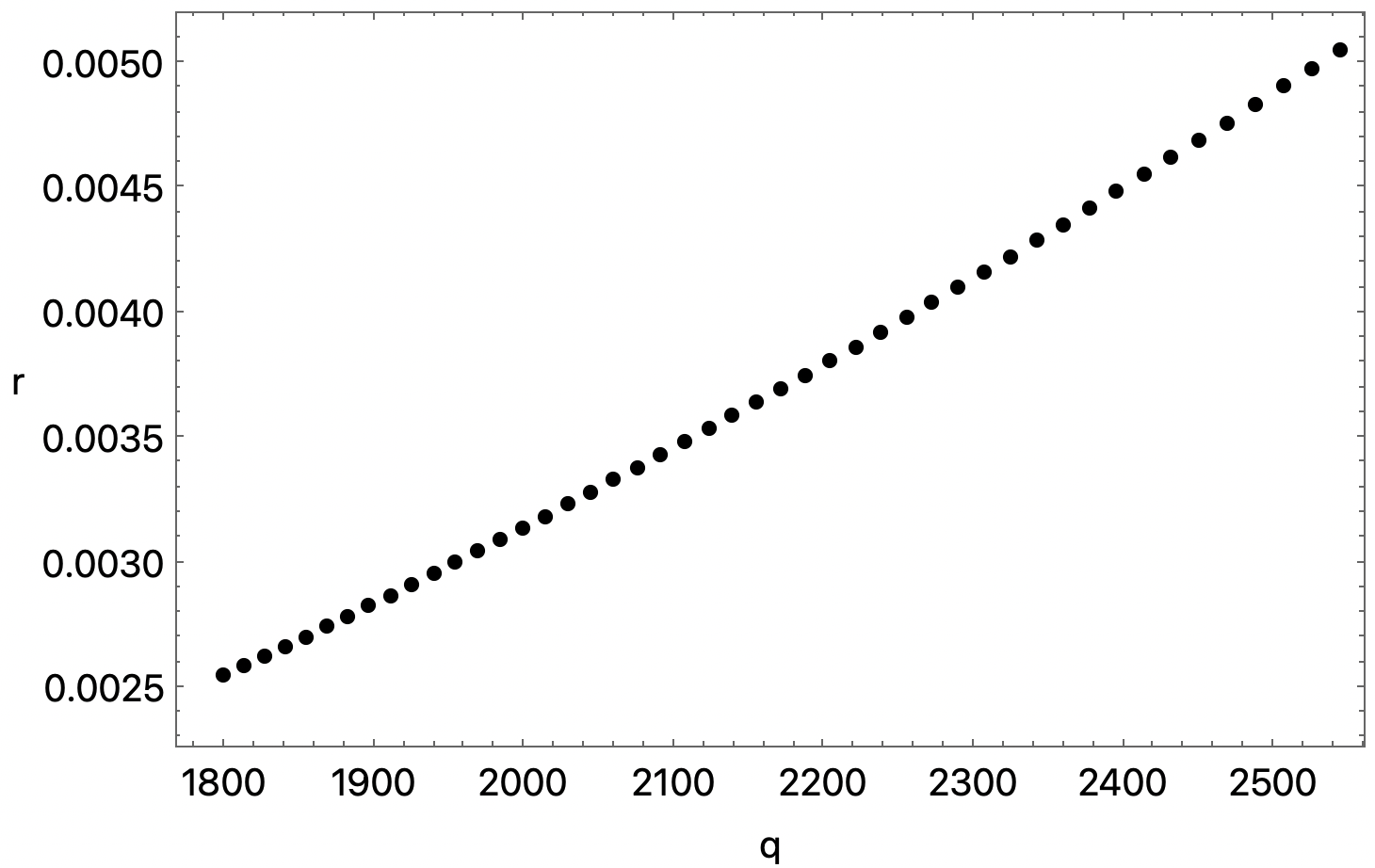}
\caption{$\tilde f=10^{-8}$}
\end{subfigure}
\caption{$n_s$ and $r$ as functions of $q$ for fixed $\tilde f=10^{-7}$
 (upper plots) and $\tilde f=10^{-8}$ (lower plots), following from the CMB
 normalization. As in Fig.~\ref{fig:ns_vs_r}, we have crossed-out the
 observationally excluded regions at $1\s$ for the spectral index~\cite
 {Planck:2018jri}, $n_s^{\rm obs}<0.9607$ and $n_s^{\rm obs}> 0.9691$. The
 tensor-to-scalar ratio, being $r\sim\mc O(10^{-3})$, is comfortably below
 the upper bound $r^{\rm obs}<0.036~{\rm at}~2\s$~\cite{BICEP:2021xfz}
 (not shown in these plots).}
\label{fig:ns_r_vs_q}
\end{figure}

The tilt $n_s$ and tensor-to-scalar ratio $r$
\be
\label{eq:infl_indexes}
n_s = 1+2\eta-6\eps \ ,~~~r = 16\eps \ ,
\ee
evaluated on~(\ref{eq:field_horizon_exit}) read
\be
\label{eq:explicit_indexes}
n_s\simeq \f{4q}{3}\f{12q \sin^2\l(\f{N}{3q}\r)-\sin\l(\f{2N}{3q}\r)}{\l(\cos\l(\f{N}{3q}\r)+4q \sin\l(\f{N}{3q}\r)\r)^2} \ ,~~~r\simeq \f{64}{9q}\f{12q \sin^2\l(\f{N}{3q}\r)-\sqrt{3}(2-\sqrt 3)\sin\l(\f{2N}{3q}\r)}{\l(8q \sin^2\l(\f{N}{3q}\r)+\sin\l(\f{2N}{3q}\r)\r)^2} \ ,
\ee 
and (for $q\sim \mc O(10^3),~N\sim \mc O(60)$) correspond to 
\be
n_s\simeq 0.9673 \ ,~~~r \simeq 0.003 \ ,
\ee 
which are fully compatible with the latest cosmological data~\cite
{Planck:2018jri,BICEP:2021xfz}. See Figs.~\ref{fig:ns_vs_r} and~\ref
{fig:ns_r_vs_q} for more precise numbers, as well as~\cite
{Salvio:2022suk,Salvio:2023dba,DiMarco:2023ncs}, where comprehensive
numerical analyses were carried out. 

Note that in the limit $q\mapsto\infty$, the tilt and tensor-to-scalar ratio
(\ref{eq:explicit_indexes}) asymptote to
\be
\label{eq:indexes}
n_s \sim 1-\f 2 N \ ,~~~r \sim \f{12}{N^2} \ , 
\ee 
meaning that they are controlled solely by inverse powers of $N$.
Interestingly, these are exactly the universal expressions for the indexes of
the Starobinsky~\cite{Starobinsky:1980te} and (metrical) Higgs
inflation~\cite{Bezrukov:2007ep}.

We close by mentioning that the implications of coupling the Weyl-invariant EC
gravity~(\ref{eq:infl_action_full}) to the Standard Model (SM) of particle
physics were studied in~\cite{Karananas:2024xja}, with an emphasis on its
finetuning issues. There, a pragmatic approach was taken by fixing the
Lorentz gauge couplings to be vanishingly small. As a result, the observed
value of the cosmological constant can be reproduced, while the
gravitationally-induced masses for the Higgs and scalar are practically zero.
Therefore, the Higgs mass is in principle computable, making the theory an
ideal playground for exploring its nonperturbative generation~\cite
{,Shaposhnikov:2018xkv,Shaposhnikov:2018jag,Karananas:2020qkp,Shaposhnikov:2020geh}.
As for the scalar, it assumes the role of the QCD axion, meaning that the
strong-CP puzzle is solved gravitationally. In our considerations here
$f,\tilde f,\tilde g$ are fixed by the primordial observables, and the
Weyl-invariant EC gravity is brought into a  domain in its parameter space
where neither the gravitational solution to the strong-CP puzzle persists,
nor the value of the cosmological constant is reproduced. It appears that if
one insists on reconciling the attractive features of the model---as far as
the SM physics is concerned---with inflation, one should give up on the
latter's geometrical origin. This leaves open the possibility that the Higgs
field be the inflaton~\cite{tbp}.

\begin{center}
\textbf{Acknowledgements} 
\end{center} 

It is a great pleasure to thank Misha Shaposhnikov and Sebastian Zell for
discussions and comments on the manuscript.

\begin{center}
\textbf{References} 
\end{center} 

{

\small

\setlength\bibsep{0pt}
    \bibliographystyle{utphys}
 \bibliography{Refs.bib}

\providecommand{\href}[2]{#2}\begingroup\raggedright\begin{thebibliography}{10}

\bibitem{Karananas:2024xja}
G.~K. Karananas, M.~Shaposhnikov, and S.~Zell, ``{Weyl-invariant
  Einstein-Cartan gravity: unifying the strong CP and hierarchy puzzles},''
  \href{http://dx.doi.org/10.1007/JHEP11(2024)146}{{\em JHEP} {\bfseries 11}
  (2024) 146}, \href{http://arxiv.org/abs/2406.11956}{{\ttfamily
  arXiv:2406.11956 [hep-th]}}.

\bibitem{Gialamas:2024iyu}
I.~D. Gialamas and K.~Tamvakis, ``{Inflation in Weyl-invariant Einstein-Cartan
  gravity},'' \href{http://arxiv.org/abs/2410.16364}{{\ttfamily
  arXiv:2410.16364 [gr-qc]}}.

\bibitem{Diakonov:2011fs}
D.~Diakonov, A.~G. Tumanov, and A.~A. Vladimirov, ``{Low-energy General
  Relativity with torsion: A Systematic derivative expansion},''
  \href{http://dx.doi.org/10.1103/PhysRevD.84.124042}{{\em Phys. Rev. D}
  {\bfseries 84} (2011) 124042},
  \href{http://arxiv.org/abs/1104.2432}{{\ttfamily arXiv:1104.2432 [hep-th]}}.

\bibitem{Rasanen:2018ihz}
S.~Rasanen, ``{Higgs inflation in the Palatini formulation with kinetic terms
  for the metric},'' \href{http://dx.doi.org/10.21105/astro.1811.09514}{{\em
  Open J. Astrophys.} {\bfseries 2} no.~1, (2019) 1},
  \href{http://arxiv.org/abs/1811.09514}{{\ttfamily arXiv:1811.09514 [gr-qc]}}.

\bibitem{Karananas:2021zkl}
G.~K. Karananas, M.~Shaposhnikov, A.~Shkerin, and S.~Zell, ``{Matter matters in
  Einstein-Cartan gravity},''
  \href{http://dx.doi.org/10.1103/PhysRevD.104.064036}{{\em Phys. Rev. D}
  {\bfseries 104} no.~6, (2021) 064036},
  \href{http://arxiv.org/abs/2106.13811}{{\ttfamily arXiv:2106.13811
  [hep-th]}}.

\bibitem{Rigouzzo:2022yan}
C.~Rigouzzo and S.~Zell, ``{Coupling metric-affine gravity to a Higgs-like
  scalar field},'' \href{http://dx.doi.org/10.1103/PhysRevD.106.024015}{{\em
  Phys. Rev. D} {\bfseries 106} no.~2, (2022) 024015},
  \href{http://arxiv.org/abs/2204.03003}{{\ttfamily arXiv:2204.03003
  [hep-th]}}.

\bibitem{Salvio:2022suk}
A.~Salvio, ``{Inflating and reheating the Universe with an independent affine
  connection},'' \href{http://dx.doi.org/10.1103/PhysRevD.106.103510}{{\em
  Phys. Rev. D} {\bfseries 106} no.~10, (2022) 103510},
  \href{http://arxiv.org/abs/2207.08830}{{\ttfamily arXiv:2207.08830
  [hep-ph]}}.

\bibitem{Salvio:2023dba}
A.~Salvio, ``{Inflation and Reheating through an Independent Affine
  Connection},'' in {\em {11th International Conference on New Frontiers in
  Physics}}.
\newblock 11, 2023.
\newblock \href{http://arxiv.org/abs/2311.02902}{{\ttfamily arXiv:2311.02902
  [hep-ph]}}.

\bibitem{DiMarco:2023ncs}
A.~Di~Marco, E.~Orazi, and G.~Pradisi, ``{Einstein\textendash{}Cartan
  pseudoscalaron inflation},''
  \href{http://dx.doi.org/10.1140/epjc/s10052-024-12482-6}{{\em Eur. Phys. J.
  C} {\bfseries 84} no.~2, (2024) 146},
  \href{http://arxiv.org/abs/2309.11345}{{\ttfamily arXiv:2309.11345
  [hep-th]}}.

\bibitem{Barker:2024dhb}
W.~Barker and S.~Zell, ``{Consistent particle physics in metric-affine gravity
  from extended projective symmetry},''
  \href{http://arxiv.org/abs/2402.14917}{{\ttfamily arXiv:2402.14917
  [hep-th]}}.

\bibitem{Gialamas:2022xtt}
I.~D. Gialamas and K.~Tamvakis, ``{Inflation in metric-affine quadratic
  gravity},'' \href{http://dx.doi.org/10.1088/1475-7516/2023/03/042}{{\em JCAP}
  {\bfseries 03} (2023) 042}, \href{http://arxiv.org/abs/2212.09896}{{\ttfamily
  arXiv:2212.09896 [gr-qc]}}.

\bibitem{Karananas:2024hoh}
G.~K. Karananas, ``{The particle content of $R^2$ gravity revisited},''
  \href{http://arxiv.org/abs/2407.09598}{{\ttfamily arXiv:2407.09598
  [hep-th]}}.

\bibitem{Karananas:2024qrz}
G.~K. Karananas, ``{The particle content of (scalar curvature)$^2$
  metric-affine gravity},'' \href{http://arxiv.org/abs/2408.16818}{{\ttfamily
  arXiv:2408.16818 [hep-th]}}.

\bibitem{Starobinsky:1980te}
A.~A. Starobinsky, ``{A New Type of Isotropic Cosmological Models Without
  Singularity},'' \href{http://dx.doi.org/10.1016/0370-2693(80)90670-X}{{\em
  Phys. Lett. B} {\bfseries 91} (1980) 99--102}.

\bibitem{Alvarez-Gaume:2015rwa}
L.~Alvarez-Gaume, A.~Kehagias, C.~Kounnas, D.~L\"ust, and A.~Riotto, ``{Aspects
  of Quadratic Gravity},'' \href{http://dx.doi.org/10.1002/prop.201500100}{{\em
  Fortsch. Phys.} {\bfseries 64} no.~2-3, (2016) 176--189},
  \href{http://arxiv.org/abs/1505.07657}{{\ttfamily arXiv:1505.07657
  [hep-th]}}.

\bibitem{Shtanov:2023lci}
Y.~Shtanov, ``{Electroweak symmetry breaking by gravity},''
  \href{http://dx.doi.org/10.1007/JHEP02(2024)221}{{\em JHEP} {\bfseries 02}
  (2024) 221}, \href{http://arxiv.org/abs/2305.17582}{{\ttfamily
  arXiv:2305.17582 [hep-ph]}}.

\bibitem{Planck:2018jri}
{\bfseries Planck} Collaboration, Y.~Akrami {\em et~al.}, ``{Planck 2018
  results. X. Constraints on inflation},''
  \href{http://dx.doi.org/10.1051/0004-6361/201833887}{{\em Astron. Astrophys.}
  {\bfseries 641} (2020) A10},
  \href{http://arxiv.org/abs/1807.06211}{{\ttfamily arXiv:1807.06211
  [astro-ph.CO]}}.

\bibitem{BICEP:2021xfz}
{\bfseries BICEP, Keck} Collaboration, P.~A.~R. Ade {\em et~al.}, ``{Improved
  Constraints on Primordial Gravitational Waves using Planck, WMAP, and
  BICEP/Keck Observations through the 2018 Observing Season},''
  \href{http://dx.doi.org/10.1103/PhysRevLett.127.151301}{{\em Phys. Rev.
  Lett.} {\bfseries 127} no.~15, (2021) 151301},
  \href{http://arxiv.org/abs/2110.00483}{{\ttfamily arXiv:2110.00483
  [astro-ph.CO]}}.

\bibitem{Bezrukov:2007ep}
F.~L. Bezrukov and M.~Shaposhnikov, ``{The Standard Model Higgs boson as the
  inflaton},'' \href{http://dx.doi.org/10.1016/j.physletb.2007.11.072}{{\em
  Phys. Lett. B} {\bfseries 659} (2008) 703--706},
  \href{http://arxiv.org/abs/0710.3755}{{\ttfamily arXiv:0710.3755 [hep-th]}}.

\bibitem{Shaposhnikov:2018xkv}
M.~Shaposhnikov and A.~Shkerin, ``{Conformal symmetry: towards the link between
  the Fermi and the Planck scales},''
  \href{http://dx.doi.org/10.1016/j.physletb.2018.06.068}{{\em Phys. Lett. B}
  {\bfseries 783} (2018) 253--262},
  \href{http://arxiv.org/abs/1803.08907}{{\ttfamily arXiv:1803.08907
  [hep-th]}}.

\bibitem{Shaposhnikov:2018jag}
M.~Shaposhnikov and A.~Shkerin, ``{Gravity, Scale Invariance and the Hierarchy
  Problem},'' \href{http://dx.doi.org/10.1007/JHEP10(2018)024}{{\em JHEP}
  {\bfseries 10} (2018) 024}, \href{http://arxiv.org/abs/1804.06376}{{\ttfamily
  arXiv:1804.06376 [hep-th]}}.

\bibitem{Karananas:2020qkp}
G.~K. Karananas, M.~Michel, and J.~Rubio, ``{One residue to rule them all:
  Electroweak symmetry breaking, inflation and field-space geometry},''
  \href{http://dx.doi.org/10.1016/j.physletb.2020.135876}{{\em Phys. Lett. B}
  {\bfseries 811} (2020) 135876},
  \href{http://arxiv.org/abs/2006.11290}{{\ttfamily arXiv:2006.11290
  [hep-th]}}.

\bibitem{Shaposhnikov:2020geh}
M.~Shaposhnikov, A.~Shkerin, and S.~Zell, ``{Standard Model Meets Gravity:
  Electroweak Symmetry Breaking and Inflation},''
  \href{http://dx.doi.org/10.1103/PhysRevD.103.033006}{{\em Phys. Rev. D}
  {\bfseries 103} no.~3, (2021) 033006},
  \href{http://arxiv.org/abs/2001.09088}{{\ttfamily arXiv:2001.09088
  [hep-th]}}.

\bibitem{tbp}
G.~K. Karananas, H.~D.~K. Nguyen, M.~Shaposhnikov, and S.~Zell. \emph{In
  preparation}.

\end{thebibliography}\endgroup
}

\end{document}